\documentclass{bmcart}

\usepackage{amsthm,amsmath,amssymb}
\usepackage[utf8]{inputenc} 

\usepackage{graphicx}

\startlocaldefs
\endlocaldefs

\begin{document}

\begin{frontmatter}

\begin{fmbox}
\dochead{Research}


\title{Propagation of quantum correlations after a quench 
in the Mott-insulator regime of the Bose-Hubbard model}


\author[
   addressref={aff1},                   
   corref={aff1},                       
   email={konstantin.krutitsky@uni-due.de}   
]{\inits{KV}\fnm{Konstantin V} \snm{Krutitsky}}
\author[
   addressref={aff1},
   email={patrick.navez@uni-due.de}
]{\inits{P}\fnm{Patrick} \snm{Navez}}
\author[
   addressref={aff1,aff2},
   email={friedemann.queisser@gmx.de}
]{\inits{F}\fnm{Friedemann} \snm{Queisser}}
\author[
   addressref={aff1},
   email={schuetz@theo.physik.uni-due.de}
]{\inits{R}\fnm{Ralf} \snm{Sch\"utzhold}}


\address[id=aff1]{
  \orgname{Fakult\"at f\"ur Physik, Universit\"at Duisburg-Essen}, 
  \street{Lotharstrasse 1},                     %
  \postcode{47057}                                
  \city{Duisburg},                              
  \cny{Germany}                                    
}
\address[id=aff2]{%
  \orgname{Department of Physics, University of British Columbia},
  \street{6224 Agricultural Road},
  \city{Vancouver}
  \postcode{V6T 1Z1},
  \cny{Canada}
}


\begin{artnotes}
\end{artnotes}

\end{fmbox}


\begin{abstractbox}

\begin{abstract} 
%
We study a quantum quench in the Bose-Hubbard model where the tunneling rate 
$J$ is suddenly switched from zero to a finite value in the Mott regime. 
In order to solve the many-body quantum dynamics far from equlibrium, we
consider the reduced density matrices for a finite number 
of lattice sites and split them up into on-site density operators, i.e., 
the mean field, plus two-point and three-point correlations etc. 
Neglecting three-point and higher correlations, we are able to numerically 
simulate the time-evolution of the on-site
density matrices and the two-point quantum correlations 
(e.g., their effective light-cone structure) for a 
comparably large number ${\cal O}(10^3)$ of lattice sites. 
%
\end{abstract}




\end{abstractbox}
%

\end{frontmatter}




\section{Introduction}

The exponential growth of the underlying Hilbert space is one of the main 
reasons for our limited ability to simulate quantum many-body systems. 
In contrast to classical many-particle systems, it is not sufficient to treat 
the state of each particle or lattice site separately -- one has to consider 
the quantum correlations (entanglement) as well.
If these quantum correlations obey certain properties, for example if they are 
sufficiently short-ranged, there are quite efficient methods for approximating 
the quantum state of the system, such as matrix-product states
\cite{Vidal04,VPC04,CV09,Sch11}.
However, in typical non-equilibrium situations
\cite{CC06,CC07,KLA07,RDYO07,RDO08,EHKKMWF09,MK09,CR10,PSSV11,R13,R14}
these quantum correlations tend to
spread
\cite{CFMSE08,FCMSE08,LK08,BPBRK12,BPCK12,NM13a,NM13b,DS13,CBSSF14,BEL14,BCKO14}
and thus are not short-ranged after some time -- which is the reason why these 
methods typically break down in such cases eventually.
In the following, we present an alternative method which fully incorporates 
quantum correlations between pairs of lattice sites at arbitrary distances. 
The price we have to pay is the neglect of genuine three-point (and higher)
correlations. 
However, comparison with exact numerical diagonalization (for small systems)
reveals that this is a reasonable approximation.
Even though the method we shall present can be applied to quite general 
lattice  systems, we are going to focus on the Bose-Hubbard model, because 
the non-equlibrium many-body quantum dynamics in this system can be tested 
in high-precision experiments using ultra-cold atoms in optical lattices
\cite{CWBD11,CBPESFGBKK12,TCFMSEB12}.

\section{Bose-Hubbard model}

We consider a system of bosons with local interactions in a lattice described 
by the Bose-Hubbard Hamiltonian
\begin{equation}
\label{Bose-Hubbard}
\hat H
=
-\frac{J}{Z}
\sum_{\mu_1\mu_2}
T_{\mu_1\mu_2}
\hat b^\dagger_{\mu_1} \hat b_{\mu_2}
+
\frac{U}{2}
\sum_{\mu} 
\hat n_{\mu}(\hat n_{\mu}-1)
\;,
\end{equation}
where $J$ is the tunneling rate for the nearest neighbors and $U>0$ is the 
interaction parameter.
The creation and annihilation operators $\hat b^\dagger_{\mu_1}$ and $\hat b_{\mu_2}$
at the lattice sites $\mu_1$ and $\mu_2$, respectively, satisfy the standard 
bosonic commutation relations, and $\hat n_{\mu}=\hat b^\dagger_\mu \hat b_\mu$ 
is the particle-number operator.
The lattice structure is encoded in the adjacency matrix $T_{\mu_1\mu_2}$ 
which equals unity if the sites labeled by $\mu_1$ and $\mu_2$ are tunneling 
neighbors (i.e., if a particle can hop from $\mu_1$ to $\mu_2$) 
and zero otherwise. 
The number of tunneling neighbors at a given site $\mu_1$ yields the 
coordination number $Z=\sum_{\mu_2} T_{\mu_1\mu_2}$. 
(We assume a translationally invariant lattice with periodic boundary 
conditions).

In order to study the time evolution of the system after a sudden change of 
the parameter $J$, we consider the density matrix describing the quantum state 
of the complete lattice which obeys the von Neumann equation ($\hbar=1$)
\begin{equation}
\label{von Neumann}
i\partial_t\hat\rho
=
\left[
    \hat H,\hat\rho
\right]
\;.
\end{equation}
Then we introduce a set of $\ell$-point reduced density matrices for the 
sites $\mu_1\mu_2\cdots\mu_\ell$ by tracing over all other sites:
\begin{equation}
\hat\rho_{\mu_1\mu_2\cdots\mu_\ell}
=
{\rm tr}_{\not\mu_1\not\mu_2\cdots\not\mu_\ell}\{\hat\rho\}
\;.
\end{equation}
These reduced density matrices can be split up into their correlated parts 
$\hat\rho_{\mu_1\mu_2\cdots\mu_\ell}^{\rm corr}$ plus all possible products of 
reduced density matrices for single lattice sites and the correlated parts 
of the reduced density matrices for smaller number of sites -- 
such that the total number of sites for the 
terms within each product equals to $\ell$.
For instance, the correlated parts of the two-point and three-point reduced 
density matrices are in anology with the cumulant expansion given by
\begin{eqnarray}
\hat\rho_{\mu_1\mu_2}^{\rm corr}
&=&
\hat\rho_{\mu_1\mu_2}-\hat\rho_{\mu_1}\hat\rho_{\mu_2}
\nonumber\\
\hat\rho_{\mu_1\mu_2\mu_3}^{\rm corr}
&=&
\hat\rho_{\mu_1\mu_2\mu_3}
-
\hat\rho_{\mu_1\mu_2}^{\rm corr}\hat\rho_{\mu_3}
-
\hat\rho_{\mu_1\mu_3}^{\rm corr}\hat\rho_{\mu_2}
-
\hat\rho_{\mu_2\mu_3}^{\rm corr}\hat\rho_{\mu_1}
-
\hat\rho_{\mu_1}\hat\rho_{\mu_2}\hat\rho_{\mu_3}
\;.
\nonumber
\end{eqnarray}
Inserting this split into the von Neumann equation~(\ref{von Neumann}) yields 
the equations of motion for the reduced density matrices $\rho_{\mu_1}$ 
for single lattice sites (i.e., the on-site matrices) and the correlated parts 
$\hat\rho_{\mu_1\mu_2\cdots\mu_\ell}^{\rm corr}$, which were already derived in 
Ref.~\cite{QKNS14}.
The equation for the on-site matrix $\hat\rho_{\mu_1}$ reads 
\begin{eqnarray}
\label{eom_rho_1}
i
\partial_t \hat \rho_{\mu_1}
&=&
\left[
\hat H_{\mu_1}
,
\hat\rho_{\mu_1}
\right]
+
\frac{1}{Z}
\sum_{\mu_2\neq\mu_1}
{\rm tr}_{\mu_2}
\left\{
\hat H_{\mu_1\mu_2} + \hat H_{\mu_2\mu_1}
,
\hat\rho_{\mu_1\mu_2}^{\rm corr} + \hat\rho_{\mu_1}\hat\rho_{\mu_2}
\right\}
\,,
\end{eqnarray}
where $\hat H_{\mu}=(U/2)\hat n_{\mu}(\hat n_{\mu}-1)$ and
$\hat H_{\mu_1\mu_2}=-J T_{\mu_1\mu_2}\hat b^\dagger_{\mu_1} \hat b_{\mu_2}$
are the local and non-local parts of the Hamiltonian.

The analogous equation for the two-point correlators  
$\hat\rho_{\mu_1\mu_2}^{\rm corr}$
contains the on-site matrices $\hat\rho_{\mu_1}$, but also the 
three-point correlations $\hat\rho_{\mu_1\mu_2\mu_3}^{\rm corr}$.
In the following, we assume that we can neglect these three-point correlations 
-- which is the main assumption of our method. 
The quality of this approximation will be discussed in the next section. 
With this assumption, we obtain the approximate equation for 
$\hat\rho_{\mu_1\mu_2}^{\rm corr}$:
\begin{eqnarray}
\label{eom_rho_2}
i
\partial_t \hat\rho_{\mu_1\mu_2}^{\rm corr}
&=&
\left[
\hat H_{\mu_1}
,
\hat\rho_{\mu_1\mu_2}^{\rm corr}
\right]
+
\frac{1}{Z}
\left[
\hat H_{\mu_1\mu_2}
,
\hat\rho_{\mu_1\mu_2}^{\rm corr}
+
\hat\rho_{\mu_1}\hat\rho_{\mu_2}
\right]
\\
&-&
\frac{1}{Z}
\hat\rho_{\mu_1}
{\rm tr}_{\mu_1}
\left\{
\left[
\hat H_{\mu_1\mu_2} + \hat H_{\mu_2\mu_1}
,
\hat\rho_{\mu_1\mu_2}^{\rm corr}
+
\hat\rho_{\mu_1}
\hat\rho_{\mu_2}
\right]
\right\}
\nonumber\\
&+&
\frac{1}{Z}
\sum_{\mu_3\neq \mu_1,\mu_2}
{\rm tr}_{\mu_3}
\left\{
\left[
\hat H_{\mu_1\mu_3} + \hat H_{\mu_3\mu_1}
,
\hat\rho_{\mu_1\mu_2}^{\rm corr}
\hat\rho_{\mu_3}
+
\hat\rho_{\mu_2\mu_3}^{\rm corr}
\hat\rho_{\mu_1}
\right]
\right\}
\nonumber\\
&+&
(\mu_1\leftrightarrow \mu_2)
\;.
\nonumber
\end{eqnarray}
As a consequence, the two equations~(\ref{eom_rho_1}) and (\ref{eom_rho_2})
form an approximate closed set of coupled non-linear equations, which 
we solve numerically. 

In the present work, we are interested in the dynamics of the system within 
the Mott-insulator phase.
Due to the fact that the $U(1)$-symmetry is not broken, $\hat\rho_\mu$ is 
diagonal in the basis of the local Fock states $|n\rangle_\mu$ and the 
equation of motion~(\ref{eom_rho_1}) simplifies in this basis to 
\begin{eqnarray}
\label{eom_O_1}
i
\partial_t
\langle \hat O_{\mu_1}^{n_1 n_1}\rangle
=
-
\frac{J}{Z}
\sum_{\mu_2}
T_{\mu_1\mu_2}
\left(
\sqrt{n_1}\,
\psi_{\mu_1\mu_2}^{n_1,n_1-1}
-
\sqrt{n_1+1}\,
\psi_{\mu_1\mu_2}^{n_1+1,n_1}
-
{\rm c.c.}
\right)
\,,
\end{eqnarray}
where $\hat O_\mu^{n_1 n_2} =| n_1\rangle_\mu\langle n_2 |$
and we use the notations
\begin{eqnarray}
\langle \hat O_\mu^{n_1 n_2}\rangle
&=&
{\rm tr} 
\left\{
    \hat\rho_{\mu}
    \hat O_{\mu}^{n_1 n_2}
\right\}
\;,
\\
\langle
\hat O_{\mu_1}^{n_1 n_2} \hat O_{\mu_2}^{n_3 n_4}
\rangle^{\rm corr}
&=&
{\rm tr}
\left\{
    \hat\rho_{\mu_1\mu_2}^{\rm corr}
    \hat O_{\mu_1}^{n_1 n_2} \hat O_{\mu_2}^{n_3 n_4}
\right\}
\;,
\nonumber\\
\psi_{\mu_1 \mu_2}^{n_1 n_2}
&=&
\langle \hat O^{n_1 n_2}_{\mu_1} \hat b_{\mu_2}\rangle^{\rm corr}
=
\sum_{n_3=0}^\infty
\sqrt{n_3+1}
\langle \hat O^{n_1 n_2}_{\mu_1} \hat O^{n_3,n_3+1}_{\mu_2} \rangle^{\rm corr}
\;.
\nonumber
\end{eqnarray}
Note that in the regime we are interested in
(vanishing order parameter: $\langle\hat b_\mu\rangle=0$),
$\langle \hat O_\mu^{n_1 n_2}\rangle$ vanishes for $n_1\ne n_2$. 
The diagonal terms $\langle \hat O_{\mu}^{n n}\rangle$ yield the probabilities 
$p_\mu(n)$ to have $n$ particles at the lattice site $\mu$.

As an additonal simplification, we neglect the term 
$[\hat H_{\mu_1\mu_2},\hat\rho_{\mu_1\mu_2}^{\rm corr}]/Z$ in Eq.~(\ref{eom_rho_2}) 
because it is of order $1/Z^2$ (see below).
As a result, the correlations 
$\langle \hat O_{\mu_1}^{n_1,n_2} \hat O_{\mu_2}^{n_3,n_4}\rangle^{\rm corr}$
vanish unless $n_1=n_2\pm1$ and $n_3=n_4\mp1$ and thus Eq.~(\ref{eom_rho_2}) simplifies to
\begin{eqnarray}
\label{eom_O_2}
&&
i\partial_t
\langle \hat O_{\mu_1}^{n_1+1,n_1} \hat O_{\mu_2}^{n_2,n_2+1}\rangle^{\rm corr}
=
    U
    \left(
        n_2-n_1
    \right)
\langle
    \hat O_{\mu_1}^{n_1+1,n_1} \hat O_{\mu_2}^{n_2,n_2+1}
\rangle^{\rm corr}
\\
&&
-
\frac{J}{Z}
T_{\mu_1\mu_2}
\sqrt{n_1+1}
\sqrt{n_2+1}
\left(
    \langle \hat O_{\mu_1}^{n_1+1,n_1+1}\rangle
    \langle \hat O_{\mu_2}^{n_2,n_2}\rangle
    -
    \langle \hat O_{\mu_1}^{n_1,n_1}\rangle
    \langle \hat O_{\mu_2}^{n_2+1,n_2+1}\rangle
\right)
\nonumber\\
&&
-
\frac{J}{Z}
\sum_{\mu_3\neq \mu_1,\mu_2}
T_{\mu_3\mu_1}
\sqrt{n_1+1}
\left(
    \langle \hat O_{\mu_1}^{n_1+1,n_1+1}\rangle
    -
    \langle \hat O_{\mu_1}^{n_1,n_1}\rangle
\right)
\left(\psi_{\mu_2\mu_3}^{n_2+1,n_2}\right)^*
\nonumber\\
&&
-
\frac{J}{Z}
\sum_{\mu_3\neq \mu_1,\mu_2}
T_{\mu_3\mu_2}
\sqrt{n_2+1}
\left(
    \langle \hat O_{\mu_2}^{n_2,n_2}\rangle
    -
    \langle \hat O_{\mu_2}^{n_2+1,n_2+1}\rangle
\right)
\psi_{\mu_1\mu_3}^{n_1+1,n_1}
\,.
\nonumber
\end{eqnarray}
We have checked numerically that neglecting or including the term 
$[\hat H_{\mu_1\mu_2},\hat\rho_{\mu_1\mu_2}^{\rm corr}]/Z$ does not produce any 
visible difference in the results presented in this work. 
For other correlation functions such as 
$\langle\hat n_{\mu_1}\hat n_{\mu_2}\rangle$,
however, the term $[\hat H_{\mu_1\mu_2},\hat\rho_{\mu_1\mu_2}^{\rm corr}]/Z$
becomes important. 

In the following we will study the dynamics of the system with one boson per 
site which is initially in the ground state of the Bose-Hubbard Hamiltonian 
at $J=0$.
In this case, $\langle \hat O_\mu^{n n}\rangle=\delta_{n,1}$ and all correlations 
$\langle \hat O_{\mu_1}^{n_1+1,n_1} \hat O_{\mu_2}^{n_2,n_2+1}\rangle^{\rm corr}$ vanish.
The system of non-linear equations~(\ref{eom_O_1}) and ~(\ref{eom_O_2}) can be 
solved using different strategies.
One possibility is to make a perturbative expansion with respect to the 
inverse coordination number $1/Z$ assuming that the reduced density matrices 
scale as~\cite{QKNS14}
\begin{eqnarray}
\label{1/Z}
\hat\rho_{\mu}
=
{\cal O}\left(1\right)
\;,\quad
\hat\rho_{\mu_1\mu_2}^{\rm corr}
=
{\cal O}
\left(
    \frac{1}{Z}
\right)
\;,\quad
\hat\rho_{\mu_1\mu_2\mu_3}^{\rm corr}
=
{\cal O}
\left(
    \frac{1}{Z^2}
\right)
\;,\quad
{\rm etc.}
\end{eqnarray}
This method allows to get approximate analytical results,
see, e.g.,~\cite{QKNS14,NS10,QNS12,QNS13,NQS14}.
However, exact solutions of the truncated set of non-linear 
equations~(\ref{eom_O_1}) and ~(\ref{eom_O_2}) 
can only be obtained numerically.
Before we start to discuss the results obtained using the two strategies,
we would like to provide a justification of the neglect of three-point 
correlations which will be done in the next section.

\section{Two-point versus three-point correlations in 1D and 2D}
\label{Two-point versus three-point}

In this section, we present exact numerical results for two-point and 
three-point correlation functions in a one-dimensional chain consisting 
of 11 lattice sites and in a two-dimensional square lattices of $3\times3$
lattice sites. 
They are obtained by full diagonalization of the Bose-Hubbard Hamiltonian 
with periodic boundary conditions without any truncation of the Hilbert space.
This allows us to calculate exactly the complete time evolution of any 
quantity as well as their mean values averaged over an infinite time.

In Fig.~\ref{Fig:corr2p3p}, we display the time evolution of the 
nearest-neighbor two-point correlations $\langle\hat b_1^\dagger\hat b_2\rangle$
as well as two exemplary three-point correlations 
$\langle \hat b_{1} (\hat b_2^\dagger)^2 \hat b_3\rangle$ 
and 
$\langle \hat n_{1} \hat b_2^\dagger \hat b_3\rangle^{\rm corr}
\equiv
\langle \hat n_{1} \hat b_2^\dagger \hat b_3\rangle
-\langle \hat n_{1}\rangle \langle \hat b_2^\dagger \hat b_3\rangle$.
By comparing these three-point correlators to others such as 
$\langle(\hat b_2^\dagger)^2\hat b_{3}\hat b_4\rangle$, we found that 
the former $\langle \hat b_{1} (\hat b_2^\dagger)^2 \hat b_3\rangle$ 
is larger than the latter 
$\langle(\hat b_2^\dagger)^2\hat b_{3}\hat b_4\rangle$
-- i.e., the quantities plotted in Fig.~\ref{Fig:corr2p3p} 
yield an upper bound. 
%
%
In one dimension, the two-point correlation 
$\langle\hat b_1^\dagger\hat b_2\rangle$ oscillates around an average value 
of about 0.2 while the largest three-point correlator 
$\langle \hat b_{1} (\hat b_2^\dagger)^2 \hat b_3\rangle$ 
always lies far below $\langle\hat b_1^\dagger\hat b_2\rangle$ 
and oscillates around an average value of roughly 0.03. 
Thus, already in one dimension, the neglect of the three-point correlations 
seems to be an approximation which is not too bad. 
As expected from the $1/Z$-arguments above, the two-point correlation 
$\langle\hat b_1^\dagger\hat b_2\rangle$ is weaker in two dimensions with 
an average value of approximately 0.12 while the three-point correlator 
$\langle \hat b_{1} (\hat b_2^\dagger)^2 \hat b_3\rangle$ 
is even more suppressed and oscillates around an average value of 
roughly 0.01. 

In summary, although the three-point correlations are not zero, 
they are indeed smaller than the two-point functions 
$\langle \hat b_1^\dagger \hat b_2\rangle$, which justifies our approximation 
up to a certain accuracy -- even in one dimension. 
Let us roughly estimate the impact of the neglected three-point correlations
on a local observable such as $\langle\hat n_\mu^2\rangle$. 
The time-derivative $\partial_t\langle\hat n_\mu^2\rangle$ yields terms 
containing two-point correlations such as 
$J\langle\hat n_\mu\hat b_\mu^\dagger\hat b_\nu\rangle$ 
which are fully included in our method. 
However, the time-derivative of that term gives contributions containing 
three-point correlations, for example 
$J^2\langle\hat b_\lambda(\hat b_\mu^\dagger)^2\hat b_\nu\rangle$, 
which are neglected within our method.  
Thus, in order to achieve an accuracy $\varepsilon$ (say, one percent)  
for the local observable $\langle\hat n_\mu^2\rangle$, we could integrate 
the evolution equation for a time $t$ until 
$(Jt)^2\langle\hat b_1(\hat b_2^\dagger)^2\hat b_3\rangle={\cal O}(\varepsilon)$ 
before the accumulated error induced by the neglect of the 
three-point correlations may become too large. 
With $J/U=0.1$, this gives a time of order $10/U$ in two dimensions and a 
somewhat shorter time in one dimension.

%

\section{\label{sec:order1}First order in $1/Z$}

In this section we discuss the approximate analytical solutions of 
Eqs.~(\ref{eom_O_1}) and~(\ref{eom_O_2}) obtained in Ref.~\cite{QKNS14} 
in first order of $1/Z$.
The probabilities of the occupation numbers larger than two vanish and
the probabilities to have zero and two particles are equal to each other. 
Their time dependence is given by
\begin{equation}
\label{quench-p0}
p_\mu(0)=p_\mu(2)
=
\frac{4J^2}{N}
\sum_\mathbf{k}
T_\mathbf{k}^2
\,
\frac{1-\cos(\omega_\mathbf{k}t)}{\omega^2_\mathbf{k}}
\;,
\end{equation}
where $\omega_\mathbf{k}$ denotes the dispersion relation of the particle-hole 
excitations
\begin{eqnarray}
\label{eigen-frequency}
\omega_\mathbf{k}
=
\sqrt{U^2-6 J UT_\mathbf{k}+J^2 T_\mathbf{k}^2}
\,,
\end{eqnarray}
and $T_\mathbf{k}$ is the Fourier transform of the adjacency matrix
\begin{equation}
T_\mathbf{k}
=
\frac{1}{NZ}
\sum_{\mu_1\mu_2}
T_{\mu_1\mu_2}
\exp
\left[
    i
    \mathbf{k}
    \cdot
    \left(
        \mathbf{x}_{\mu_1}-\mathbf{x}_{\mu_2}
    \right)
\right]
\,.
\end{equation}
For hypercubic lattices in $D$ dimensions (with $Z=2D$), we have 
\begin{equation}
T_\mathbf{k}
=
\frac{1}{D}
\sum_{d=1}^D
\cos k_d
\;,\quad
k_d \in \frac{2\pi}{L_d}\mathbb{Z}
\;,
\end{equation}
where $L_d$ is the size of the lattice along the direction $d$.
The summation over ${\bf k}$ in Eq.~(\ref{quench-p0}) is restricted to 
the first Brillouin zone.

The time evolution of the one-body density matrix is described by 
the equation
\begin{equation}
\label{quench-b+b}
\langle\hat{b}^\dagger_{\mu_1}
\hat{b}_{\mu_2}^{} \rangle
=
\frac{4JU}{N}
\sum_\mathbf{k}
T_\mathbf{k}\,
\frac{1-\cos(\omega_\mathbf{k}t)}{\omega^2_\mathbf{k}}
\,
\exp
\left[
    i
    \mathbf{k}
    \cdot
    \left(
        \mathbf{x}_{\mu_1}-\mathbf{x}_{\mu_2}
    \right)
\right]
\,.
\end{equation}
For large distances $\mathbf{x}_{\mu_1}-\mathbf{x}_{\mu_2}$, we may approximate 
the $\mathbf{k}$-summation/integration via the saddle-point method. 
Then, for a given direction in ${\bf k}$-space, the maximum group velocity
${\bf v}_{\rm max}={\rm max}\;\nabla_{\bf k}\omega_{\bf k}$
determines the maximum propagation speed of correlations,
i.e., the effective light cone. 
In a hypercubic lattice in $D$ dimensions with small $J$, for example,
it is given by $v_{\rm max}\approx 3J/D$ along the lattice axes and by 
$v_{\rm max}\approx 3J/\sqrt{D}$ along the diagonal 
(where all the components of ${\bf v}_{\rm max}$ are equal to each other).
A similar result has been obtained in Ref.~\cite{BPCK12} for the 
one-dimensional Bose-Hubbard model.
For an experimental realization, see, e.g., Ref.~\cite{CBPESFGBKK12}.

The above approximate analytical solution (\ref{quench-p0}) for the 
probability of zero occupation number $p_\mu(0)$  is compared in 
Fig.~\ref{Fig:an-small} with the results obtained by exact diagonalization 
for one-dimensional and two-dimensional lattices -- which was already 
presented in Ref.~\cite{QKNS14}.
Although Eq.~(\ref{quench-p0}) predicts the correct behavior
for short times $t={\cal O}(1/U)$,
the discrepancy with exact numerical data on a longer time scale
is quite noticeable, especially in one dimension. 
The same feature was also observed for the two-point correlation function 
$\langle\hat{b}^\dagger_{\mu_1}\hat{b}_{\mu_2}^{} \rangle$.
As we will see in the next section, a significant improvement can be achieved 
if the calculations are done in a non-perturbative manner. 

\section{Numerical solution of coupled equations}

Now we turn to the numerical solution of the coupled set of 
equations~(\ref{eom_O_1}) and~(\ref{eom_O_2}).
In order to check the quality of the obtained results, we do calculations 
first for small systems, where we can compare with exact diagonalization. 
The data presented in Fig.~\ref{Fig:num-small} indicate that there is 
certainly an improvement compared to the perturbative solutions discussed 
in section~\ref{sec:order1}.
The major difference between the two methods is the following:
The perturbative approach discussed in section~\ref{sec:order1}
is based on a linearization around the Gutzwiller solution 
$\hat\rho_\mu^0=|1\rangle_\mu\langle1|$ which facilitates (approximate)
analytical solutions -- whereas coupled equations~(\ref{eom_O_1}),~(\ref{eom_O_2})
fully include the evolution
of $\hat\rho_\mu(t)$ and its back-reaction onto the dynamics of the correlations. 

As can be observed in Fig.~\ref{Fig:num-small}, we obtain a quantitative 
agreement not just for short times $t={\cal O}(1/U)$, but also for 
intermediate times $t={\cal O}(10/U)$ in one dimension and 
$t={\cal O}(30/U)$ in two dimensions.
These time scales are consistent with the estimate discussed at the end of 
section~\ref{Two-point versus three-point} because the data in 
Fig.~\ref{Fig:num-small} are directly related to the quantity
$\langle\hat n_\mu^2\rangle\approx p_\mu(1)+4p_\mu(2)\approx 1+2p_\mu(0)$
considered there due to $p_\mu(0)\approx p_\mu(2)\gg p_\mu(3)$.
Moreover, even for longer times $t={\cal O}(100/U)$, we find that our method 
reproduces qualitative features reasonably well, especially in two dimensions. 
There seems to be a shift of the time coordinate of a few percent 
(for certain modes),  
which might be explainable by an effective renormalization of $J$ 
due to the neglected higher-order contributions as discussed in 
Ref.~\cite{QKNS14}.  

Having found that the neglect of three-point correlations yields quantitative 
agreement for short and intermediate times and reproduces qualitative features 
for longer time scales, we now apply the same method to larger lattices for 
long times (which are hardly reachable by other methods). 
From a minimal standpoint, this is a study of what happens if these three-point
correlations are neglected. 
However, in view of the agreement observed above and since the three-point 
correlators in large lattices are presumably comparable to those in 
Fig.~\ref{Fig:corr2p3p}, we expect that this procedure again yields 
quantitative agreement for short and intermediate times and reproduces 
qualitative features for longer time scales. 

\subsection{Dynamics after quench in a large one-dimensional chain}

The time dependence of the probabilities $p_\mu(n)$ as well as the one-body 
density matrix $\langle\hat{b}^\dagger_{\mu_1}\hat{b}_{\mu_2}^{} \rangle$
for a chain of $50$ sites is shown in Fig.~\ref{Fig:num-large1D}.
The probabilities $p_\mu(0)$ and $p_\mu(2)$ almost coincide and higher 
occupations $p_\mu(n)$ with $n\ge3$ are negligible.

After several oscillations at the initial stage, $p_\mu(n)$ stabilizes -- 
which is often referred to as prethermalization~\cite{BBW04,KWE11,KISD11}. 
However, at a later time $t>150/U$, we see a revival of oscillations 
(followed again by stabilization). 
Comparison with the dynamics of the correlations displayed in the lower panel 
of Fig.~\ref{Fig:num-large1D} reveals that this revival time roughly coincides 
with the time the correlations need to move through the whole chain. 
(Note that this picture is symmetric with respect to $s=25$ due to the imposed 
periodic boundary conditions.)   

This coincidence can be explained via the following intuitive quasi-particle 
picture:
The quantum quench generates correlated particle-hole pairs 
(analogous to cosmological particle creation, for example~\cite{BD84,SU13}) 
which move away in opposite directions due to quasi-momentum conservation.
Initially, the wave-functions of these particles and holes have a large 
overlap and thus their phase coherence (i.e., their correlation) affects the 
local probabilities $p_\mu(0)$ and $p_\mu(2)$ leading to an oscillation 
(destructive versus constructive interference).
When they move away from each other, this overlap decreases and thus the 
local probabilities $p_\mu(0)$ and $p_\mu(2)$ settle down to a quasi-stationary 
value (loss of phase coherence). 
However, when these correlated particles and holes meet again 
(after one round trip), their wave-functions overlap once more and thus 
the oscillation of $p_\mu(n)$ induced by (destructive or constructive) 
interference is repeated. 
Therefore, these revivals can be regarded as a finite-size effect 
induced by the periodic boundary conditions: enlarging the system size shifts 
these revivals to later times.

Consistent with this quasi-particle picture, one can clearly see a propagating 
front of correlations moving with a constant velocity $v_{\rm max}$, which 
corresponds to the time-dependent distance between the correlated particles 
and holes. 
This velocity $v_{\rm max}$ agrees quite well with the analytical result 
obtained in the first order of the $1/Z$ expansion, see the previous section.



\subsection{Dynamics after quench in a large two-dimensional 
square lattice}

Numerical results for a two-dimensional lattice of $30\times30$ sites are 
presented in Figs.~\ref{Fig:aacorr2D} and~\ref{Fig:num-large2D}.
The spatial dependences of the function 
$\langle\hat b^\dagger_{\mu_1} \hat b_{\mu_2}^{\phantom{\dagger}}\rangle$
at different moments of time are shown in Fig.~\ref{Fig:aacorr2D},
where each panel extends over the whole lattice and the reference site 
labeled by $\mu_1$ is in the middle of the panels.
The four-fold symmetry of these images reflects the lattice geometry.
The propagation is anisotropic with the velocities being maximal
along the lattice diagonals and minimal along the axes. 
Similar results were recently obtained by the variational Monte Carlo 
method~\cite{CBSSF14}. 

Fig.~\ref{Fig:num-large2D} shows the time evolution of the probabilities 
$p_\mu(n)$ as well as of the correlation function
$\langle\hat b^\dagger_{\mu_1} \hat b_{\mu_2}^{\phantom{\dagger}}\rangle$
along a lattice axis (middle panel) and along a diagonal (lower panel).
Note that the distances in the latter plot are rescaled by a factor of 
$\sqrt{2}$ in comparison to the former one.

Consistent with the $1/Z$ expansion, the overall amplitude of $p_\mu(n)$
and $\langle\hat b^\dagger_{\mu_1} \hat b_{\mu_2}^{\phantom{\dagger}}\rangle$ is 
reduced by roughly a factor of $1/2$. 
The propagation velocity of the wave front is also in a good agreement with 
the analytical predictions of the $1/Z$ expansion.

As in the one-dimensional setup, we see again a revival of oscillations of 
$p_\mu(n)$ after the propagation of the correlations through the whole lattice.
However, this revival effect is not so pronounced as in one dimension.
We attribute this reduction to the enlarged phase space of the quasi-particles
in two dimensions, which results in a weaker phase coherence.  
Intuitively speaking, the reunions of particle-hole pairs with different 
momenta do not occur at the same time. 

\section{Conclusions}

We have developed a method which allows us to simulate the dynamics of 
quantum correlations in lattices of comparably large sizes for relatively 
long time scales.
The formalism is based on the equations of motion~(\ref{eom_rho_1}) and 
(\ref{eom_rho_2}) for the reduced density matrices for one and two lattice 
sites etc. 
This infinite set of equations is truncated such that the on-site density 
matrices $\rho_{\mu_1}$ and two-point correlations $\hat\rho_{\mu_1\mu_2}^{\rm corr}$
are taken into account exactly whereas three-point and higher correlations 
are neglected. 

This approximation is motivated by a perturbative expansion in powers of 
the inverse coordination number $1/Z$.
As a result, our method is very suited for high-dimensional lattices -- 
which are quite hard to treat within most other approaches (such as matrix 
product states).
However, exact diagonalization for small lattices in one and two 
dimensions (see Fig.~\ref{Fig:corr2p3p}) shows that the underlying 
approximation is not too bad in this regime as well:
The three-point correlations such as 
$\langle \hat b_{1} (\hat b_2^\dagger)^2 \hat b_3\rangle$ 
are much smaller than the two-point function 
$\langle\hat b_1^\dagger\hat b_2\rangle$, see  Fig.~\ref{Fig:corr2p3p},
while other three-point correlations such as 
$\langle \hat n_{1} \hat b_2^\dagger \hat b_3\rangle^{\rm corr}
\equiv
\langle \hat n_{1} \hat b_2^\dagger \hat b_3\rangle
-\langle \hat n_{1}\rangle \langle \hat b_2^\dagger \hat b_3\rangle$
are even more suppressed. 
This observation is partly explained by the fact that we are still 
comparably deep in the Mott phase. 
Using strong-coupling perturbation theory type arguments, the two-point 
function $\langle\hat b_1^\dagger\hat b_2\rangle$ scales with the first 
order in $J/U$ while the three-point correlations are second-order effects. 
Note, however, that we do not use any expansion in $J/U$ -- our only 
approximation is the neglect of the three-point and higher correlations.

Comparison to exact diagonalization for small lattices 
(in one and two dimensions, see Fig.~\ref{Fig:num-small}) shows that our 
truncation scheme yields quantitative agreement for short $t={\cal O}(1/U)$
and intermediate $t={\cal O}(10/U)$ times and reproduces qualitative features 
for longer time scales $t={\cal O}(100/U)$.
Since the three-point correlations in large lattices are presumambly 
comparable to those in small lattices (depicted in Fig.~\ref{Fig:corr2p3p}),
we expect that this agreement applies to larger lattices as well.


As an application, we study the dynamics of interacting bosons in a lattice 
as described by the Bose-Hubbard Hamiltonian~(\ref{Bose-Hubbard}) 
after quench within the Mott-insulator regime.
We find that the time evolution of the local particle-number distribution
$p_\mu(n)$ is directly related to the propagation of the correlations
$\langle\hat b_1^\dagger\hat b_2\rangle$.  
In particular, we find the revival of oscillations of the local quantities 
$p_\mu(n)$ after the correlations traverse the whole lattice. 
In two dimensions, the propagation of correlations is anisotropic with the 
velocity being minimal along the lattices axes and maximal along the 
diagonals.
 
Of course, we do not claim that our method is generally superior to other 
approaches such as matrix-product states (MPS) or the density-matrix 
renormalization group (DMRG). 
As always, different approaches have their own advantages and drawbacks.
Apart from the aforementioned applicability to higher-dimensional lattices
and the ability to treat long-range correlations, 
the advantages of our method are the following:
Most importantly, the computational complexity of our method scales 
polynomially (instead of exponentially) with system size -- even in the 
presence of long-range correlations. 
Note that this is different for matrix-product states, for example, 
where the required internal matrix dimension scales exponentially with 
the entanglement entropy which is proportional to the system size in 
these non-equilibrium situations~\cite{Sch11}.
Furthermore, our method can be applied to general lattice structures with 
periodic or open boundary conditions and does not contain any free fit 
parameter. 
In contrast to weak-coupling methods (see, e.g.,~\cite{EHKKMWF09}), it is 
particularly suited for the regime of strong interactions $U$.

As an outlook, our method can be extended easily to inhomogeneous lattices 
with manageable computational overhead -- which facilitates taking into 
account the trap potential as well as disorder potentials.
Even though the initial state used in this work had no correlations, it is 
trivial to incorporate initial correlations (such as a thermal state with 
finite $J$, for example).
In order to improve the accuracy, it is also possible to shift the truncation 
by taking into account three-point correlations 
(see the Note added below),
either fully or only within a finite range -- or to approximate them by 
suitable functions of the on-site density matrices $\rho_{\mu_1}$ and two-point 
correlations $\hat\rho_{\mu_1\mu_2}^{\rm corr}$. 
Finally, our method can be applied to other systems such as spin lattices or 
the Fermi-Hubbard model, for instance.  
In this context, it would be interesting to study the relation between our 
method and time-dependent (i.e., non-equilibrium) dynamical mean-field theory 
(t-DMFT). 
This approach is based on the approximation of the self-energy by a local 
quantity which becomes exact in the limit $Z\to\infty$.
Thus, it displays some similarities to the $1/Z$-expansion in Eq.~(\ref{1/Z}) 
but with the important difference that the hopping term in 
Eq.~(\ref{Bose-Hubbard}) scales with $1/Z$ instead of the 
$1/\sqrt{Z}$-scaling used in (fermionic) t-DMFT. 
As a result, the limit $Z\to\infty$ in t-DMFT becomes more complicated than 
in our method where all the correlations decay as a power of $1/Z$, see 
Eq.~(\ref{1/Z}).  
Nevertheless, it would be interesting to compare our method to t-DMFT, 
which should be the subject of further studies.

\section{Appendix}

Since we approximate the full von Neumann equation~(\ref{von Neumann}) within 
our truncation scheme (by omitting three-point and higher correlations), it is 
a good idea to ask the question of whether this scheme breaks some of the 
important conservation laws of the underlying system -- which could then lead 
to unphysical results. 

First we check that our truncated set of equations~(\ref{eom_O_1}) 
and~(\ref{eom_O_2}) conserves the trace of the density operator 
${\rm tr}\{\hat\rho\}=1$ (conservation of probability). 
For the reduced density matrices, this implies 
${\rm tr}\{\hat\rho_\mu\}=1$ and 
${\rm tr}\{\hat\rho_{\mu_1\mu_2}^{\rm corr}\}=0$.
By taking the trace of equations~(\ref{eom_rho_1}) and (\ref{eom_rho_2}),
we see that these traces are conserved exactly within our truncated scheme.
However, one should be aware that the positivity of $\hat\rho_\mu$ is not 
guaranteed.
Even though this was not a problem for our simulations, we encountered 
numerical instabilities related to this issue in some other extremal cases.   

By multiplying Eq.~(\ref{eom_rho_1}) with $\hat n_{\mu_1}$, taking the trace, 
and summing over $\mu_1$, we find that the total number of particles is also 
conserved
\begin{eqnarray}
\partial_t\langle \hat N\rangle
=
\sum_{\mu}\partial_t\langle \hat n_{\mu}\rangle
=0
\,.
\end{eqnarray}
Similarly, one can show that the variance $\langle \hat N^2\rangle$ of the 
total particle number is also conserved exactly by the 
equations~(\ref{eom_rho_1}) and (\ref{eom_rho_2}). 

Finally, equations~(\ref{eom_O_1}) and~(\ref{eom_O_2}) exactly conserve the 
total energy which is given by
\begin{eqnarray}
E
=
\langle \hat H\rangle
=
-
\frac{J}{Z}
\sum_{\mu_1\mu_2}
T_{\mu_1\mu_2}
\langle \hat b_{\mu_1}^\dagger \hat b_{\mu_2}^{\phantom{\dagger}}\rangle
+
\frac{U}{2}
\sum_\mu
\sum_{n=0}^\infty
n(n-1)
p_\mu(n)
\;,
\end{eqnarray}
where
\begin{equation}
\langle
    \hat b^\dagger_{\mu_1}
    \hat b^{\phantom{\dagger}}_{\mu_2}
\rangle
\equiv
{\rm tr}
\left\{
    \hat\rho_{\mu_1\mu_2}^{\rm corr}\,
    \hat b^\dagger_{\mu_1}
    \hat b^{\phantom{\dagger}}_{\mu_2}
\right\}
=
\sum_{n=0}^\infty
\sqrt{n+1}\,
\psi_{\mu_1\mu_2}^{n+1,n}
\end{equation}
are the entries of the one-body density matrix.

\section{Note added}

After sumbitting our manuscript, we completed the first run of simulations 
which fully include all three-point correlations (in addition to the on-site 
density matrices and two-point correlations) but neglect four-point and 
higher correlators. 
Even though these results are still preliminary, they indicate that the 
accuracy and reliability for longer time scales can be improved significantly, see the plots
for small lattices presented in Fig.~\ref{Fig:3p} and compare with those in Fig.~\ref{Fig:num-small}.
The same calculations for larger lattices lead to results which are very similar to those 
presented in Figs.~\ref{Fig:num-large1D},~\ref{Fig:aacorr2D},~\ref{Fig:num-large2D}.
More details will be published elsewhere.

\begin{backmatter}

\section*{Competing interests}
The authors declare that they have no competing interests.

\section*{Author's contributions}
All four authors participated in the design of the research, 
analysis of the results, and writing of the paper.

\section*{Acknowledgements}
The authors acknowledge valuable discussions with W.~Hofstetter, S.~Kehrein, 
C.~Kollath, A.~Rosch, M.~Vojta and many others.   
F.Q. is supported by the Templeton foundation (grant number JTF 36838). 
This work was supported by the DFG (SFB-TR12). 


\bibliographystyle{bmc-mathphys} 




\section*{Figures}

\begin{figure}[h!]
\centering
\includegraphics[width=0.9\textwidth]{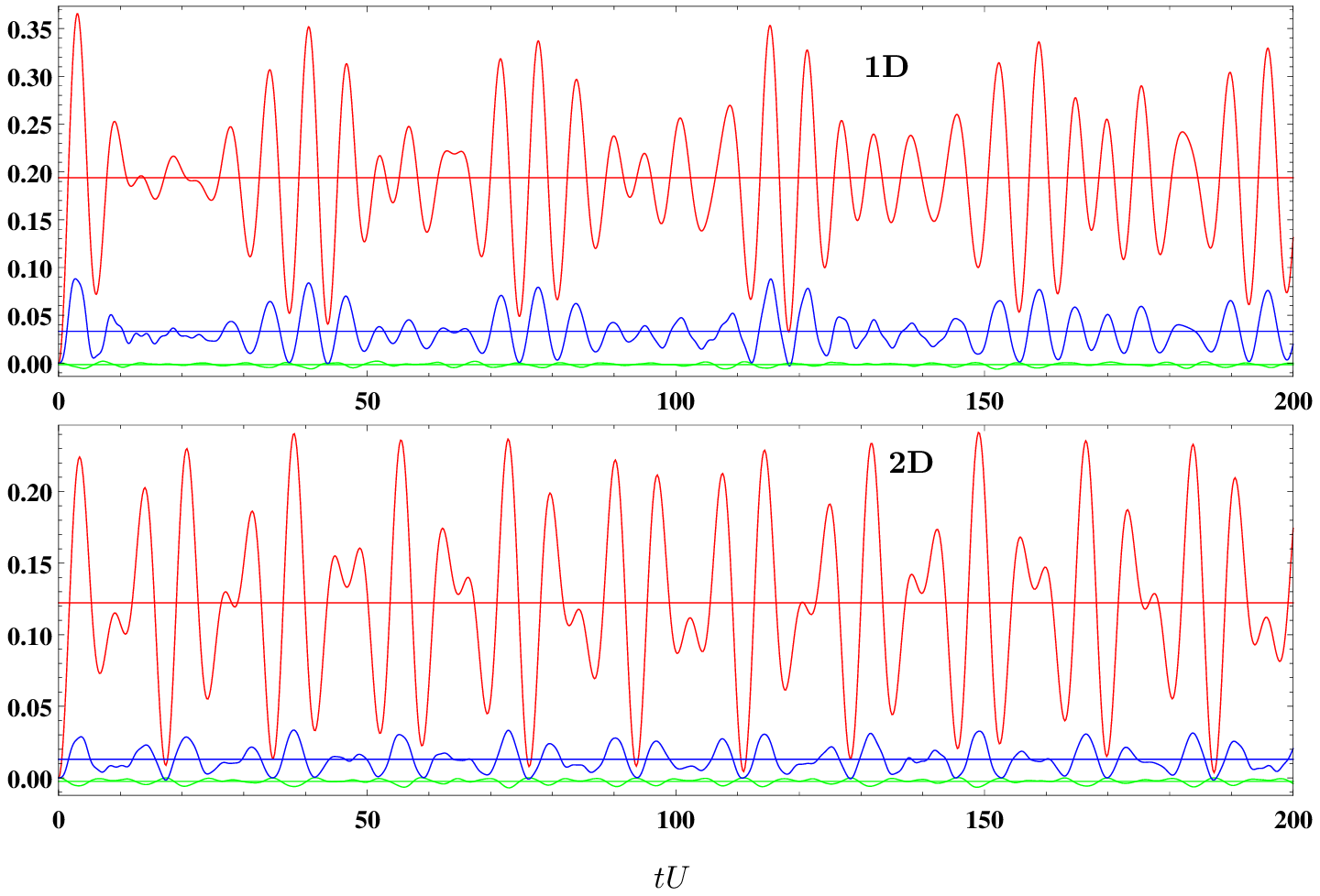}
\caption
{
\csentence{Two-point and three-point correlations.}
Correlation functions
$\langle \hat b_1^\dagger \hat b_2\rangle$~(red),
$\langle \hat b_{1} (\hat b_2^\dagger)^2 \hat b_3\rangle$~(blue),
$\langle \hat n_{1} \hat b_2^\dagger \hat b_3\rangle^{\rm corr}$~(green)
after quench $J/U=0\to0.1$ calculated by exact diagonalization for one-dimensional lattice of $11$ sites~{\bf(1D)}
and for two-dimensional lattice of $3\times3$ sites~{\bf(2D)}.
Straight horizontal lines show the values averaged over an infinite evolution time.
}
\label{Fig:corr2p3p}
\end{figure}

\begin{figure}[h!]
\centering
\includegraphics[width=0.9\textwidth]{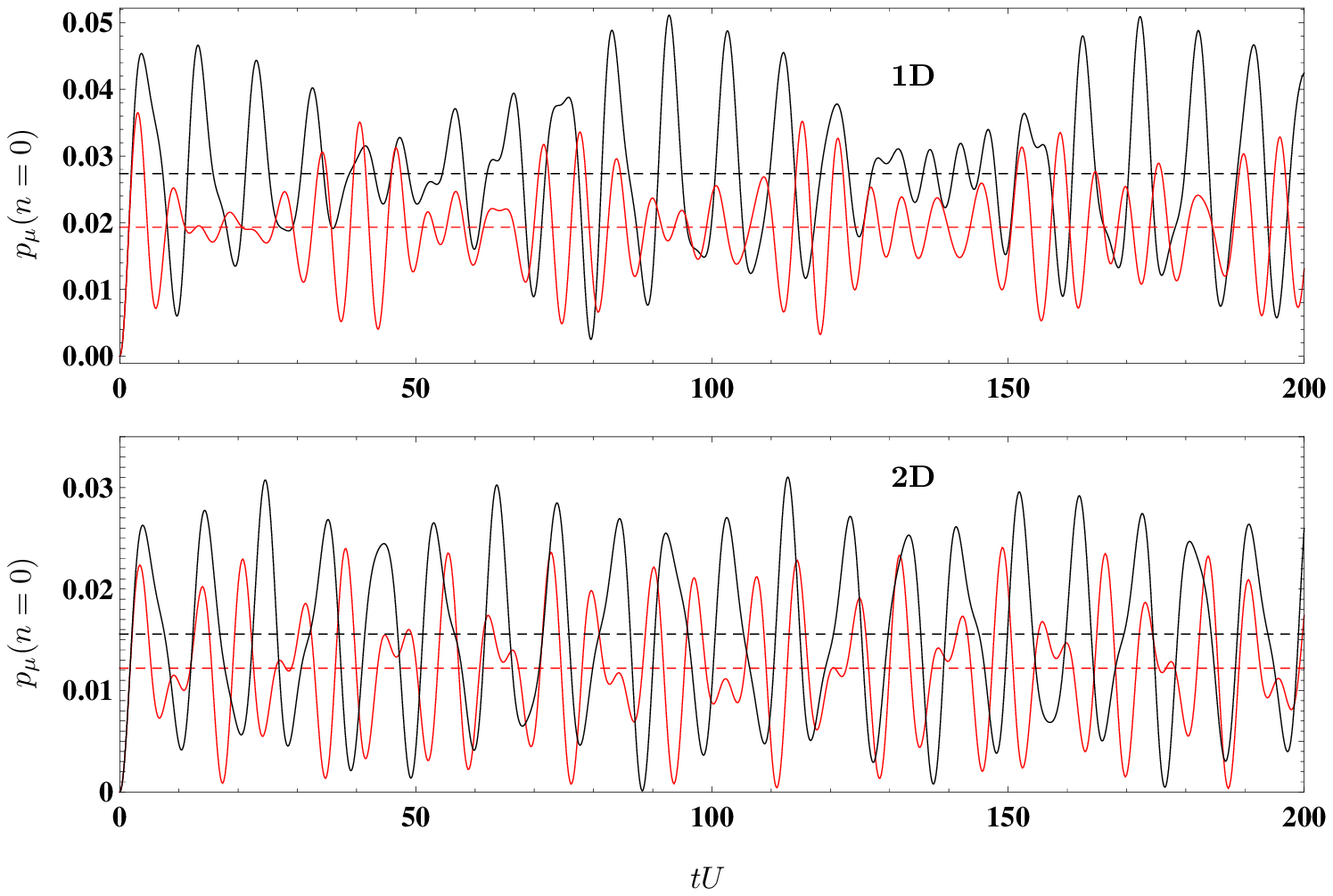}
\caption
{
\csentence{Approximate analytical solutions.}
Probability to have no particles on a lattice site in one-dimensional lattice of $11$ sites~({\bf1D})
and for two-dimensional lattice of $3\times3$ sites~({\bf2D}).
Red lines -- exact diagonalization, black lines -- Eq.~(\ref{quench-p0}).
Dashed horizontal lines show the values averaged over an infinite evolution time.
}
\label{Fig:an-small}
\end{figure}

\begin{figure}[h!]
\centering
\includegraphics[width=0.9\textwidth]{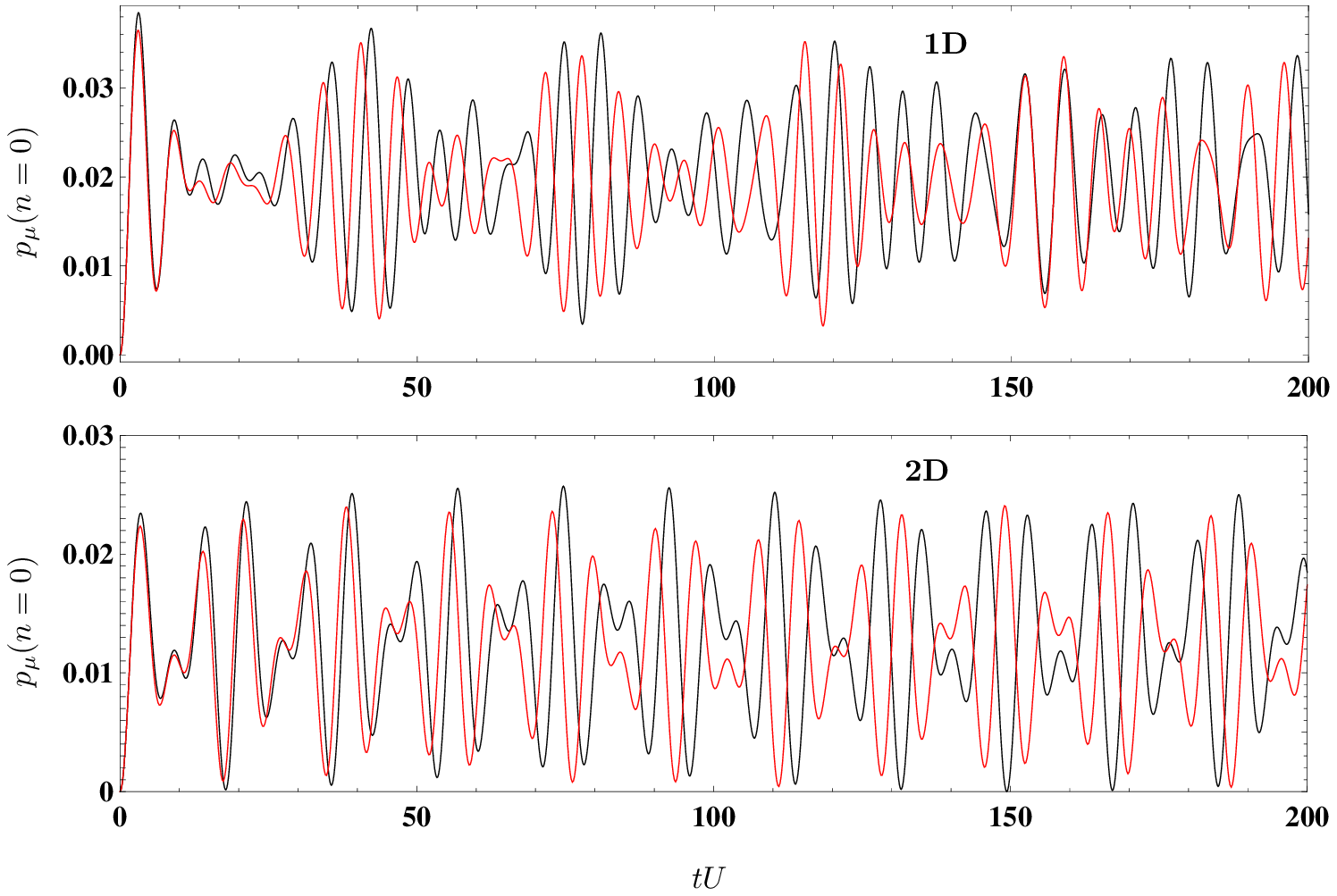}
\caption
{
\csentence{Numerical solutions for small lattices.}
Probability to have no particles on a lattice site in one-dimensional lattice of $11$ sites~({\bf1D})
and for two-dimensional lattice of $3\times3$ sites~({\bf2D}).
Red lines -- exact diagonalization (the same as in Fig.~\ref{Fig:an-small}),
black lines -- numerical solution of Eqs.~(\ref{eom_O_1}),~(\ref{eom_O_2}).
}
\label{Fig:num-small}
\end{figure}

\begin{figure}[h!]
\centering
\includegraphics[width=0.9\textwidth]{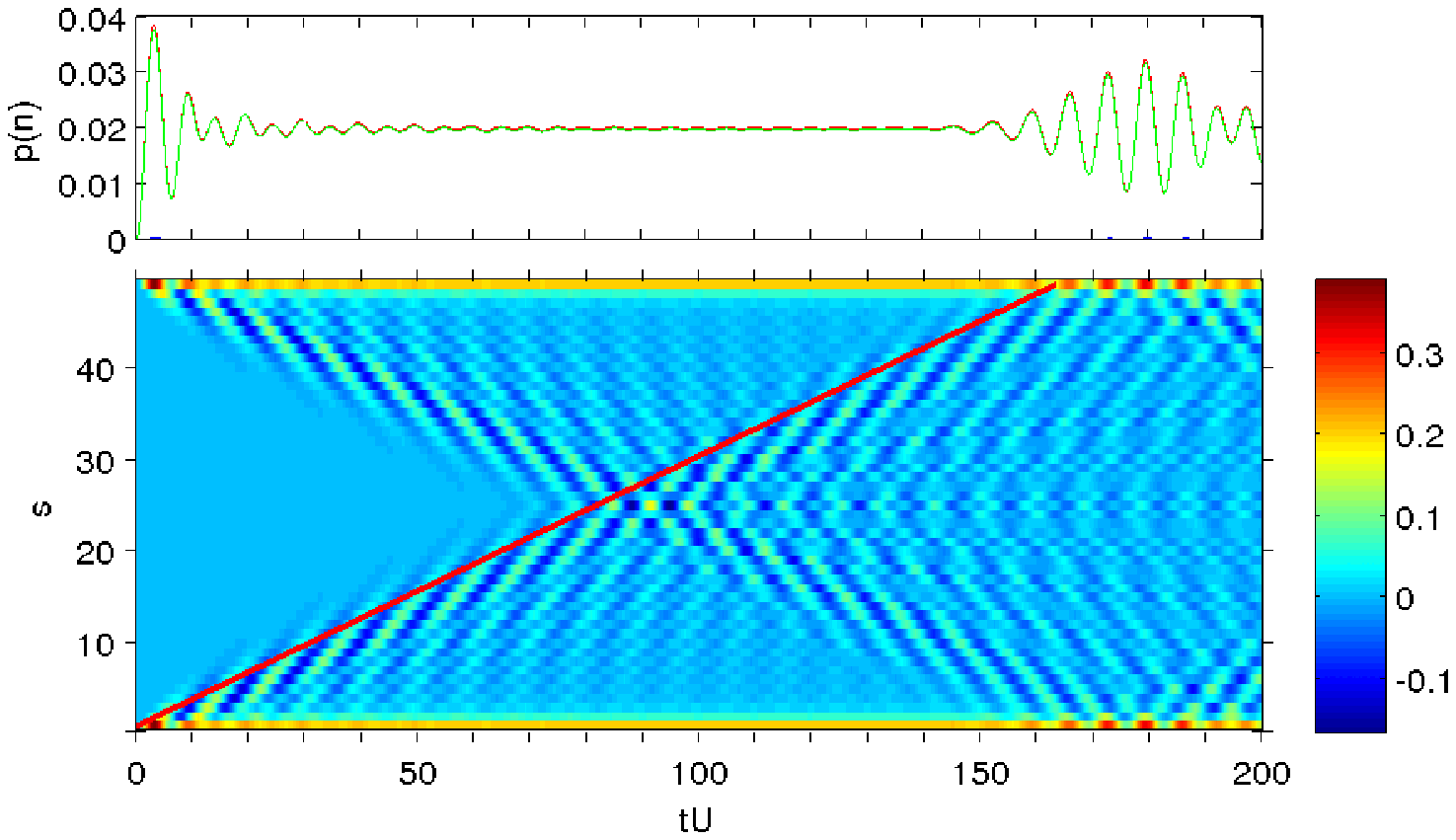}
\caption
{
\csentence{Numerical solutions for large lattice in 1D.}
Calculations for $50$ sites.
Upper panel:
Probabilities to have zero~(red) and two~(green) particles on a lattice site.
Lower panel:
$\langle\hat b_\mu^\dagger \hat b_{\mu+s}\rangle$, $s=1,\dots,49$.
Red line $s=v_{\rm max}t$, $v_{max}=3J$.
}
\label{Fig:num-large1D}
\end{figure}

\begin{figure}[h!]
\centering
\includegraphics[width=0.9\textwidth]{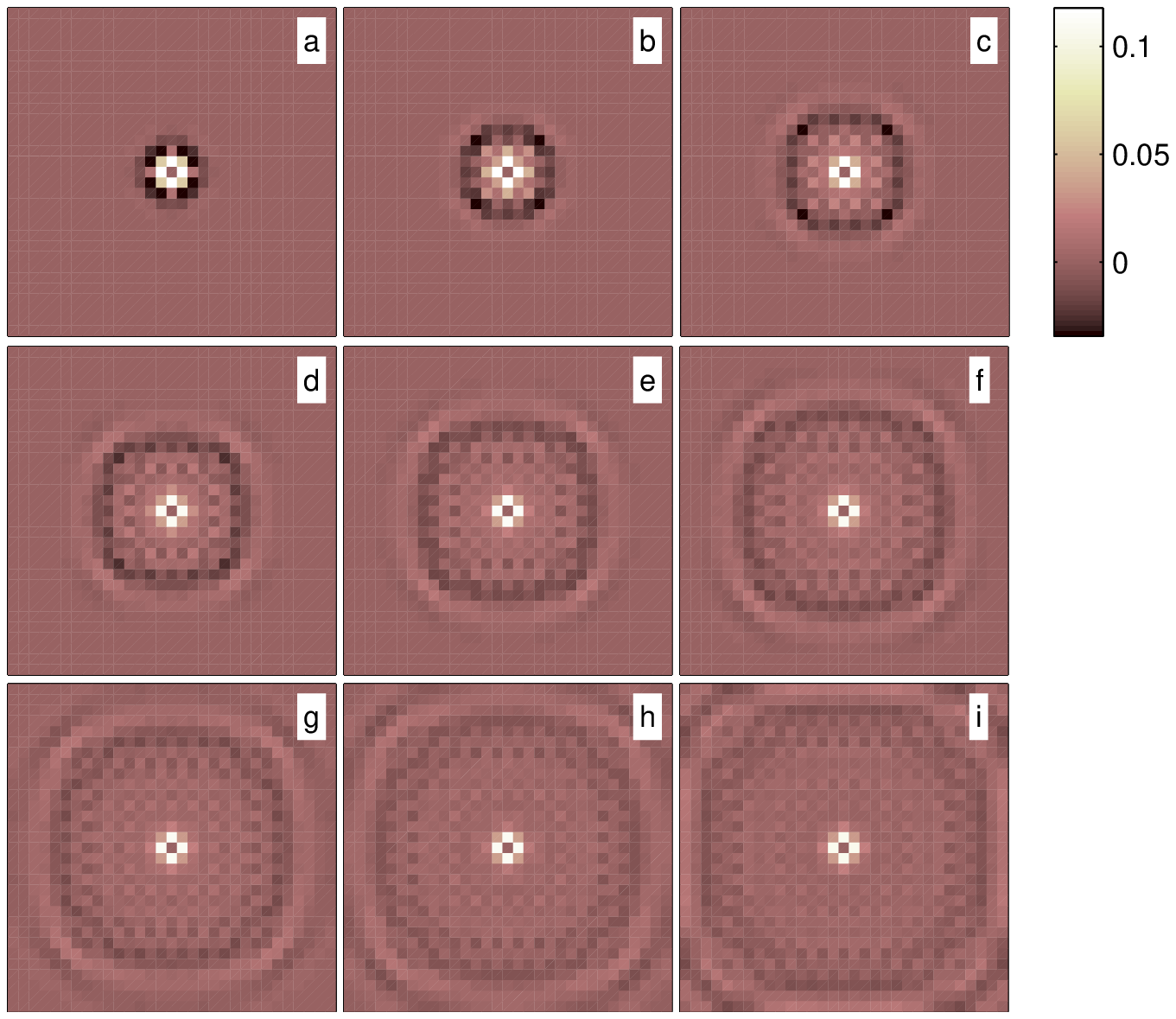}
\caption
{
\csentence{Anisotropic propagation of correlations in 2D.}
Calculations for $30\times30$ sites.
Coordinate dependences of the correlation function $\langle\hat b_{\mu_1}^\dagger \hat b_{\mu_2}\rangle$
at different times after quench:
$tU=10$~(a), $20$~(b), $30$~(c), $40$~(d), $50$~(e), $60$~(f), $70$~(g), $80$~(h), $90$~(i).
}
\label{Fig:aacorr2D}
\end{figure}

\begin{figure}[h!]
\centering
\includegraphics[width=0.9\textwidth]{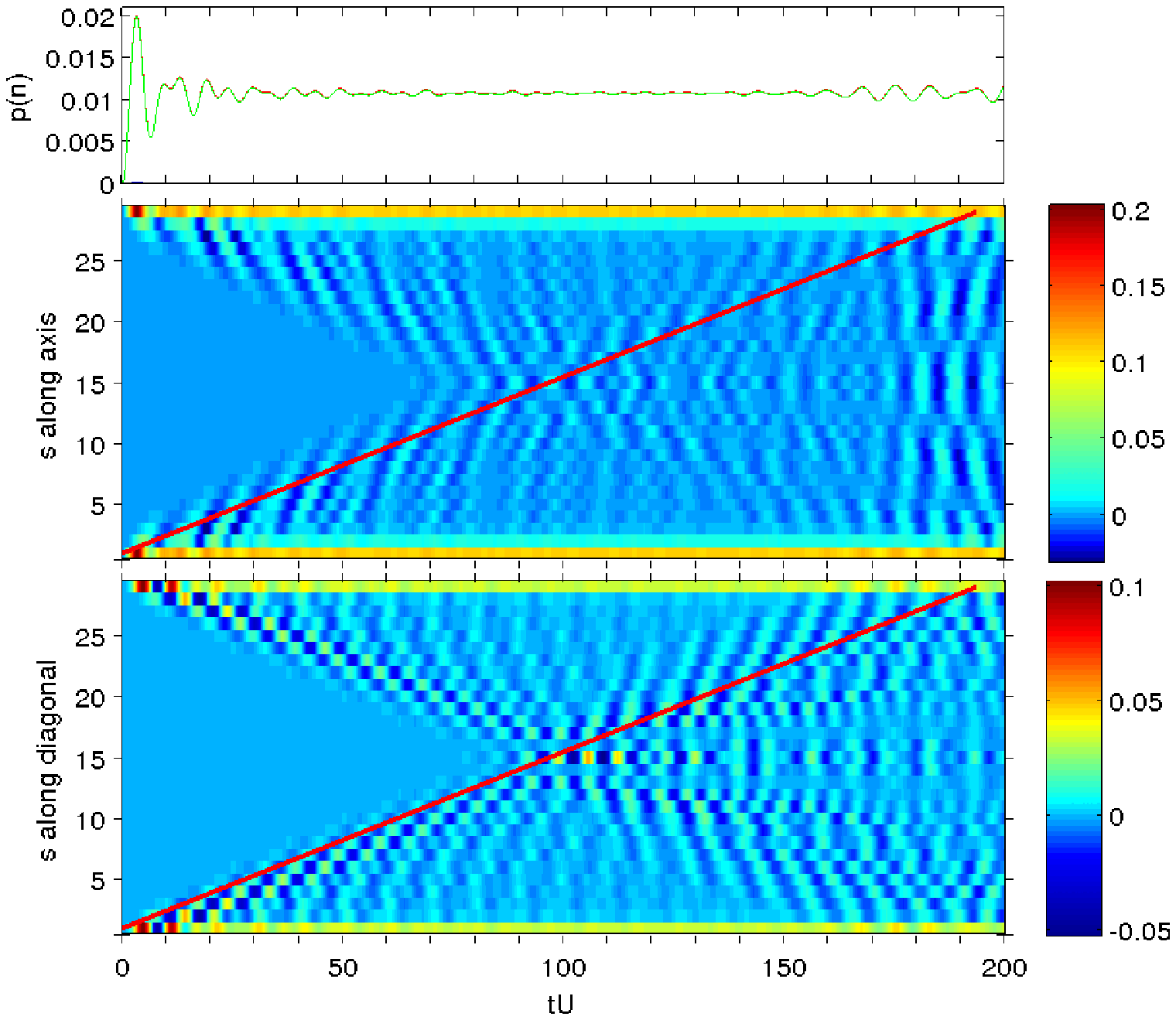}
\caption
{
\csentence{Numerical solutions for large lattice in 2D.}
Calculations for $30\times30$ sites.
Upper panel:
Probabilities to have zero~(red) and two~(green) particles on a lattice site.
Middle panel:
$\langle\hat b_\mu^\dagger \hat b_{\mu+{\bf e}_1 s}\rangle$, $s=1,\dots,29$.
Red line $s=v_{\rm max}t$, $v_{\rm max}=3J$.
Lower panel:
$\langle\hat b_\mu^\dagger \hat b_{\mu+({\bf e}_1+{\bf e}_2)s}\rangle$, $s=1,\dots,29$.
Red line $s=\sqrt{2}\,v_{\rm max}t$, $v_{\rm max}=3J/\sqrt{2}$.
${\bf e}_1$ and ${\bf e}_2$ are unit vectors along the lattice axes.
}
\label{Fig:num-large2D}
\end{figure}

\begin{figure}[h!]
\centering
\includegraphics[width=0.9\textwidth]{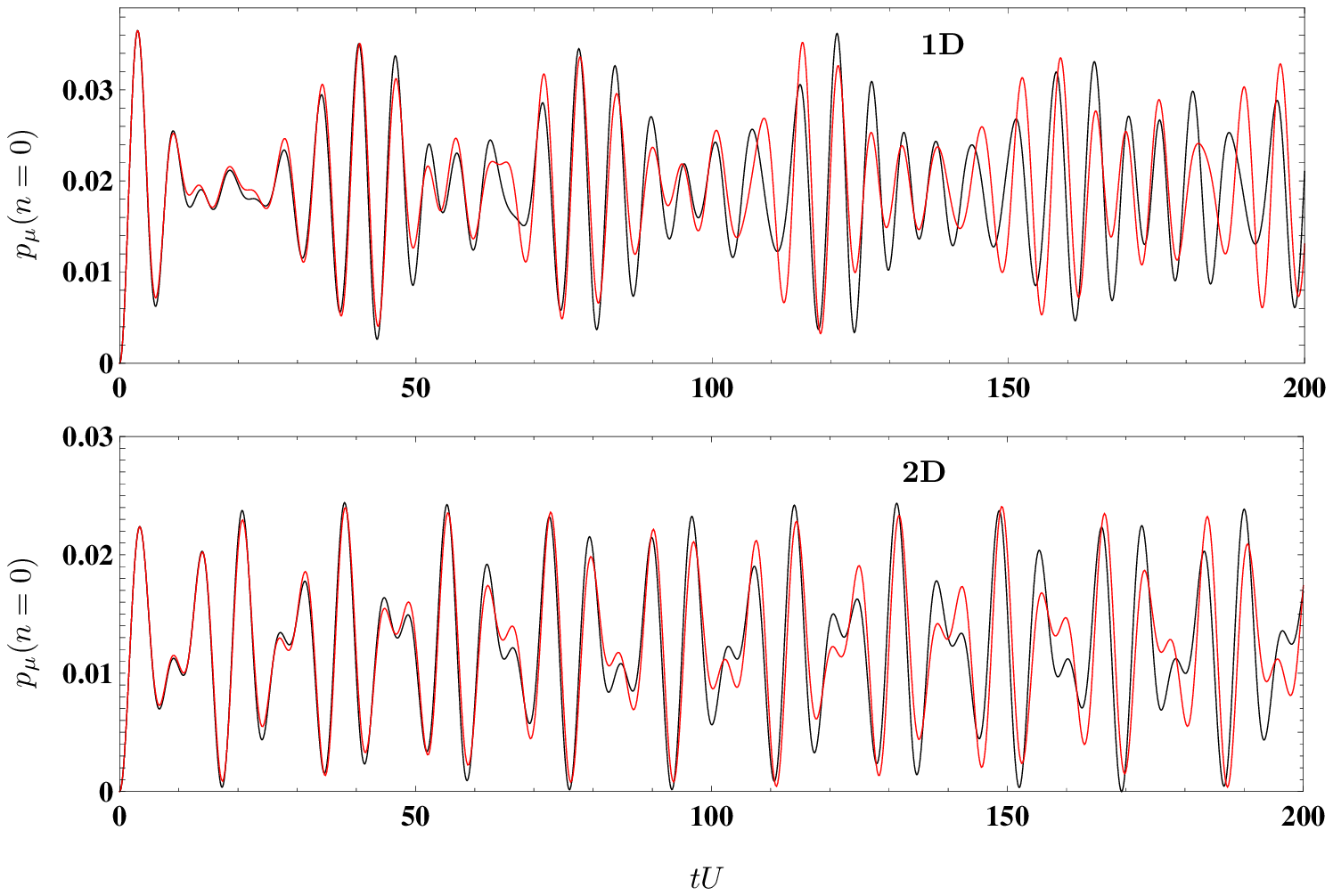}
\caption
{
\csentence{Numerical solutions for small lattices: three-point correlations.}
Probability to have no particles on a lattice site in one-dimensional lattice of $11$ sites~({\bf1D})
and for two-dimensional lattice of $3\times3$ sites~({\bf2D}).
Red lines -- exact diagonalization (the same as in Fig.~\ref{Fig:an-small}),
black lines -- numerical solution of the equations of motion for the reduced density matrices
including three-point correlations.
}
\label{Fig:3p}
\end{figure}

\section*{Additional Files}
  \subsection*{Additional file 1 --- fig1.eps}

  \subsection*{Additional file 2 --- fig2.eps}

  \subsection*{Additional file 3 --- fig3.eps}

  \subsection*{Additional file 4 --- fig4.eps}

  \subsection*{Additional file 5 --- fig5.eps}
  
  \subsection*{Additional file 6 --- fig6.eps}

  \subsection*{Additional file 7 --- fig7.eps}

\end{backmatter}
\end{document}